\documentclass[iop,revtex4]{emulateapj}
\usepackage{lscape} 
\usepackage{natbib}

\usepackage{rotating}
\usepackage{amsmath}
\newcommand \kms{km~$\rm{s}^{-1}$}

\newfont{\rten}{cmr10} 
\def \arcdeg{\hbox{$^\circ$}}

\begin{document}

\title{Could 1I/'Oumuamua be an icy fractal aggregate?}

\author{Amaya Moro-Mart\'{\i}n$^{1}$}

\altaffiltext{1}{Space Telescope Science Institute, 3700 San Martin Drive, Baltimore, MD 21218, USA; amaya@stsci.edu}

\begin{abstract}
1I/'Oumuamua is the first interstellar interloper to be detected, and it shows a non-gravitational acceleration that cannot be accounted for by outgassing, given the strict upper limits of outgassing evident from {\it Spitzer} observations, unless the relative abundances of the common volatiles are very different to those in comets. As an alternative, it has been suggested that its peculiar acceleration is due to radiation pressure, requiring a planar-sheet geometry of an unknown natural or artificial origin. Here we assess whether or not the internal structure of 1I/'Oumuamua, rather than its geometry, could support a radiation-pressure-driven scenario. We adopt a mass fractal structure and find that the type of aggregate that could yield the required area-to-mass ratio would have to be extraordinarily porous, with a density $\sim$ 10$^{-5}$ g cm$^{-3}$. Such porous aggregates can naturally arise from the collisional grow of icy dust particles beyond the snowline of a protoplanetary disk, and we propose that 1I/'Oumuamua might be a member of this population. This is a hypothesis worth investigating because, if this were the case, 1I/'Oumuamua would have opened a new observation window on to the study of the building blocks of planets around other stars. This could set unprecedented constraints on planet formation models.\end{abstract}


\section{Introduction}
\label{Introduction}

1I/'Oumuamua is the first interstellar interloper to be detected (Williams \citeyear{Williams2017}). Even though it was the subject of an intense observational campaign, its brief visit left several key questions unanswered. One of them is the number density of free-floating 1I/'Oumuamua-like objects implied from the inferred detection frequency, an aspect that relates to 1I/'Oumuamua's origin. Based on expectations from planetesimal formation models and observations, studies prior to its detection had estimated that the number density of ejected free-floating planetesimals would be so low that the detection of one of these objects crossing the solar system would require the deep surveys of the Large Synoptic Survey Telescope (LSST) era and beyond (Moro-Mart\'{\i}n et al. \citeyear{2009ApJ...704..733M}). Therefore, even though the detection of interlopers had been anticipated for decades, the first detection with PanSTARRS came as a surprise. Based on a careful calculation of the PanSTARRS detection volume, Do et al. (\citeyear{2018ApJ...855L..10D}) estimated the cumulative number density of 1I/'Oumuamua-like objects, assuming a 3.5 yr survey lifetime and that the object is representative of an isotropic background population. In Moro-Mart\'{\i}n (\citeyear{2018ApJ...866..131M}, \citeyear{2019AJ}), we found that their estimate is orders of magnitude larger than what would be expected from the ejection of planetesimals from circumstellar and circumbinary disks, and from the ejection of exo-Oort cloud objects under the effect of post-main-sequence mass loss and stellar encounters, even when considering the large uncertainties involved in our calculation (like the size distribution of ejected bodies). Other authors have reached a similar conclusion: the inferred number density is significantly higher than what would be expected in the context of a range of plausible origins (Gaidos et al. \citeyear{2017RNAAS...1a..13G}; Laughlin \& Batygin \citeyear{2017RNAAS...1a..43L}; Trilling et al., \citeyear{2017ApJ...850L..38T}; Do et al. \citeyear{2018ApJ...855L..10D}; Feng \& Jones \citeyear{2018ApJ...852L..27F}; Portegies Zwart et al. \citeyear{2018MNRAS.479L..17P}; Rafikov \citeyear{2018ApJ...861...35R}; Raymond et al. \citeyear{2018MNRAS.476.3031R}, \citeyear{2018ApJ...856L...7R}). This comparison is even less favorable when considering 1I/'Oumuamua's incoming velocity, found to be within 3--10~\kms~of the velocity of the local standard of rest  (Gaidos et al. \citeyear{2017RNAAS...1a..13G}; Mamajek \citeyear{2017RNAAS...1a..21M}), as it only favors parent systems with low dispersion velocities. In Moro-Mart\'{\i}n (\citeyear{2018ApJ...866..131M}, \citeyear{2019AJ}), we argued that this large discrepancy in the number density likely indicates that 1I/'Oumuamua is not representative of an isotropically distributed population, favoring the scenario that it originated from a nearby young system (as suggested by its kinematics; Gaidos et al. \citeyear{2017RNAAS...1a..13G}). 

Another open question is its physical properties (size, shape, rotational state, and albedo). Observations with the {\it Spitzer~Space~Telescope} could not detect its thermal emission, with the 3$\sigma$ upper limit at 4.5 $\mu$m leading to an effective spherical radius of less than [49, 70, 220] m and albedo greater than [0.2, 0.1, 0.01] (Trilling et al. \citeyear{2018AJ....156..261T}). Other estimates from visible/near-infrared observations lead to an effective radius in the range of 55--130 m (Jewitt et al. \citeyear{2017ApJ...850L..36J}, Banninster et al. \citeyear{2017ApJ...851L..38B}, Meech et al. \citeyear{2017Natur.552..378M}, Drahus et al. \citeyear{2018NatAs...2..407D}, Bolin et al. \citeyear{2018ApJ...852L...2B}, Frasser et al. \citeyear{2018NatAs...2..383F}), the uncertainties arising from its unknown shape and albedo. Based on its 2--2.5 mag variability (reduced to 1.5--1.9 when correcting for the phase angle of the observations; McNeill et al. \citeyear{2018ApJ...857L...1M}), it is estimated that 1I/'Oumuamua has an axis ratio ranging from 3 to 10, most likely in the 6$\pm$1:1 range (McNeill et al. \citeyear{2018ApJ...857L...1M}). This would make it unusually elongated compared to solar system objects, 
while other authors suggest it could be unusually oblate (Belton et al. \citeyear{2018ApJ...856L..21B}). 

The most recent puzzle regarding 1I/'Oumuamua is the 30$\sigma$ detection of a non-gravitational acceleration observed in its outbound orbit, showing a dependency of $\Delta a \propto \left({ r \over {\rm AU}} \right)^{n}$, with the best fit for $n$ = -2, which has been interpreted as evidence of outgassing (Micheli et al. \citeyear{2018Natur.559..223M}). Because no gas or dust has been observed in 1I/'Oumuamua (Jewitt et al. \citeyear{2017ApJ...850L..36J}, Meech et al. \citeyear{2017Natur.552..378M}), Micheli et al. (\citeyear{2018Natur.559..223M}) attributed the lack of activity to the presence of a thin insulating mantle. The newly released {\it Spitzer} results by Trilling et al. (\citeyear{2018AJ....156..261T}), however, imply strict upper limits to the CO and CO$_2$ outgassing that result in overall outgassing levels significantly lower than previously allowed by other observations. Specifically, the  {\it Spitzer} CO  3-$\sigma$ upper limit is four orders of magnitude lower than that invoked by Micheli et al. (\citeyear{2018Natur.559..223M}) to account for the observed non-gravitational acceleration, challenging the outgassing scenario. Trilling et al. (\citeyear{2018AJ....156..261T}) have suggested the existence of outgassing from a different gas species (unconstrained by {\it Spitzer}) and have put forward the proposal that it might be H$_2$O. However, under this scenario, even when adopting outgassing levels of CO and CO$_2$ at the 3-$\sigma$ upper limit, if one assumes that these species have a similar relative abundance with respect to H$_2$O as found in comets, 
the inferred level of H$_2$O outgassing would only be 1\% of that required, implying that for this scenario to work, the object would have to be devolatized of CO and CO$_2$ prior to {\it Spitzer} observations (Trilling et al. \citeyear{2018AJ....156..261T}). Another challenge is the expectation that the implied outgassing torques would have spun-up the object in a timescale of few days, leading to its breakup (Rafikov \citeyear{2018ApJ...867L..17R}). 

Given the challenges to the outgassing scenario, Bialy \& Loeb (\citeyear{2018ApJ...868L...1B}) have suggested that the non-gravitational acceleration could be due to radiation pressure, $P = C_R {L\odot \over 4 \pi r^2 c}$ (where $C_R$ is of order unity and depends on the objects composition and geometry, and $r$ is the distance to the Sun), that would produce an acceleration of 
$a = {PA \over m} = \left({L\odot \over 4 \pi r^2 c}\right) \left({A \over m}\right) C_R = 4.6 \times 10^{-5} \left({r \over {\rm AU}} \right)^{-2} \left({m/A \over {\rm g~cm^{-2}}}\right)^{-1} C_R ~{\rm cm~s}^{-2}$ (where $A$ and $m$ are the area and mass of the object, respectively). This acceleration has the same radial dependency as that found by Micheli et al. (\citeyear{2018Natur.559..223M}) as the best fit for 1I/'Oumuamua's excess acceleration,  $\Delta a = a_0 \left( r \over {\rm AU} \right)^n$, with $n$ = -2, and $a_0 = (4.92 \pm 0.16) \times 10^{-4}~{\rm cm~s}^{-2}$.
By equating the two expression for the excess acceleration, Bialy \& Loeb (\citeyear{2018ApJ...868L...1B}) found that for radiation pressure to be responsible for the observed non-gravitational acceleration, the required area-to-mass ratio of 1I/'Oumuamua would have to be 
\begin{equation}
\begin{split}
{A \over m} = {1 \over (9.3 \pm 0.3) \times 10^{-2}~C_R} ~{\rm cm^{2} g^{-1}}.
\end{split}
\label{m_a_michelli}
\end{equation}
They argued that for this condition to be fulfilled, the required physical properties would have to be that of a sheet 0.3--0.9 mm in width, suggesting 1I/'Oumuamua represents a new class of thin interstellar material of an unknown natural or artificial origin (like a lightsail). As an alternative to this planar sheet scenario, and given the strict {\it Spitzer} upper limits to outgassing, in this paper we assess whether a naturally-produced mass fractal structure with a high area-to-mass ratio could contribute to a radiation-pressure-driven 1I/'Oumuamua.   

\section{General properties of a mass fractal structure}
\label{mass_fractal}

Fractal structures exhibit self-similar and scale-invariant properties at all levels of magnification. They are found in many forms of nature and are thought to arise because their formation processes involve an element of stochasticity, like particle collisions in a solution, in a turbulent circumstellar cloud, or in a protoplanetary disk, environments characterized by low local particle concentrations and large diffusion lengths. Interplanetary dust particles show a fractal structure and laboratory studies of analogs of their cores (unaffected by atmospheric entry or flash heating from solar flares) indicate that these aggregates are mass fractal with a mass fractal dimension of ${N_f}$ = 1.75 (Katyal et al. \citeyear{2014JQSRT.146..290K}). 

For a mass fractal composed of self-similar structures, the number of primary particles, $N$, scales as $N = \left({D \over D_0}\right)^{N_f}$, where $D_0$ is the primary particle size (assuming a single-size distribution), $D$ is the aggregate size, and ${N_f}$ is the mass fractal dimension, an index that characterizes its space-filling capacity, i.e. the packing efficient of the aggregate. ${N_f}$ is in the range from 1--3, with ${N_f} \sim$ 3 corresponding to compact, dense structures and ${N_f} \sim$ 1 corresponding to "stringy" ones. The mass of the fractal, $m$, scales with the number of primary particles, $m \propto N \propto \left( {D \over D_0} \right)^{N_f}$, while its volume, $V$, is $V \propto D^3$. The resulting bulk density, $\rho$, scales as $\rho = \left({ m \over V }\right) \propto D^{N_f-3}$, i.e. it decreases as the aggregate size increases. 

Following Bowers et al. (\citeyear{bowers17}), that studied mass fractals in the context of marine flocs and their interaction with light, the area-to-mass ratio of a mass fractal is given by ${A \over m}$, where the area is $A = D^2$ (i.e. not assuming that these irregular fractal structures have spherical symmetry), and the mass is $m = N \rho_0 D_0^3 = \left({D \over D_0}\right)^{N_f} \rho_0 D_0^3$, where $\rho_0$ is the primary particle bulk density, leading to

\begin{equation}
\begin{split}
{A \over m} = {D^2 \over \left({D \over D_0}\right)^{N_f} \rho_0 D_0^3} = \left({D_0 \over D}\right)^{N_f-2} {1 \over \rho_0 D_0}.
\end{split}
\label{a_m_fractal}
\end{equation}
The area-to-mass ratio decreases as $N_f$ increases because, as the structure becomes more compact, the primary particles hide behind each other. In the limit of  compact structures, $N_f = 3$, the area-to-mass ratio becomes ${A \over m} =  {1 \over \rho_0 D}$, as expected for an object with uniform bulk density. For $N_f = 2$, the area-to-mass ratio, ${A \over m} =  {1 \over \rho_0 D_0}$, becomes independent of the size $D$, as it can be seen in Figure \ref{figure1}. Equation \eqref{a_m_fractal} is not valid in the $N_f = 1$ limit because, as Bowers et al. (\citeyear{bowers17}) point out, for these tenuous and stringy mass fractals, $A \propto D$  [instead of $A = D^2$  assumed in the derivation of Equation \eqref{a_m_fractal}], which would lead to ${A \over m} \propto  {1 \over \rho_0 D_0^2}$. 

The bulk density of a mass fractal is given by 
\begin{equation}
\begin{split}
\rho = {m \over V } = {\left({D \over D_0}\right)^{N_f} \rho_0 D_0^3 \over D^3}= \left({D \over D_0}\right)^{N_f-3}\rho_0,
\end{split}
\label{rho_fractal}
\end{equation}

and following Richardson et al. (\citeyear{2002aste.book..501R}), its porosity would be
  
\begin{equation}
\begin{split}
Porosity = 1 - {{\rm sum~of~primary~particle~volumes \over bulk~volume}} =\\
 1 - {N V_0 \over V} = 1 - {{\left(D \over D_0 \right)^{N_f}} D_0^3 \over D^3} = 1 - \left({D \over D_0}\right)^{N_f-3}.
\end{split}
\label{porosity_fractal}
\end{equation}

\subsection{Application to 1I/'Oumuamua}
\label{mass_fractal_1I}

Regarding the primary particle size, $D_{\rm 0}$, we consider the range 0.1 $\mu$m to 1 m ($D_0$ = 10$^{-5}$ cm, 10$^{-4}$ cm, 10$^{-3}$ cm, 10$^{-2}$ cm, 10$^{-1}$ cm, 1 cm, 10 cm, and 100 cm), assuming a single-size distribution and a bulk density of $\rho_{\rm 0}$ = 1 g/cm$^3$.  Regarding the overall aggregate size, $D$, we adopt $D$ = 98 m, 140 m, and 440 m, corresponding to the upper limits derived by Trilling et al. (\citeyear{2018AJ....156..261T}), for albedos of 0.2, 0.1, and  0.01, respectively\footnote{These values are in broad in agreement with other estimates of 1I/'Oumuamua's area, $A$, based on its observed magnitude: $A \approx 8\cdot10^{6}~\alpha^{-1}$ cm, where $\alpha$ is the albedo (Jewitt et al. \citeyear{2017ApJ...850L..36J}). Assuming $A = D^2$, this leads to $D \approx (8\cdot10^{6} \alpha^{-1})^{1/2}$ cm and using this expression, $\alpha$ = 0.04 (Fitzsimmons et al. \citeyear{2018NatAs...2..133F}), $\alpha$ = 0.1 (Jewitt et al. \citeyear{2017ApJ...850L..36J}), and $\alpha$ = 0.14  (Do et al. \citeyear{2018ApJ...855L..10D}) would lead to diameters of $D$ = 141 m, 89 m, and 76 m, respectively.}.

\begin{figure}
\begin{center}
\includegraphics[width=8cm]{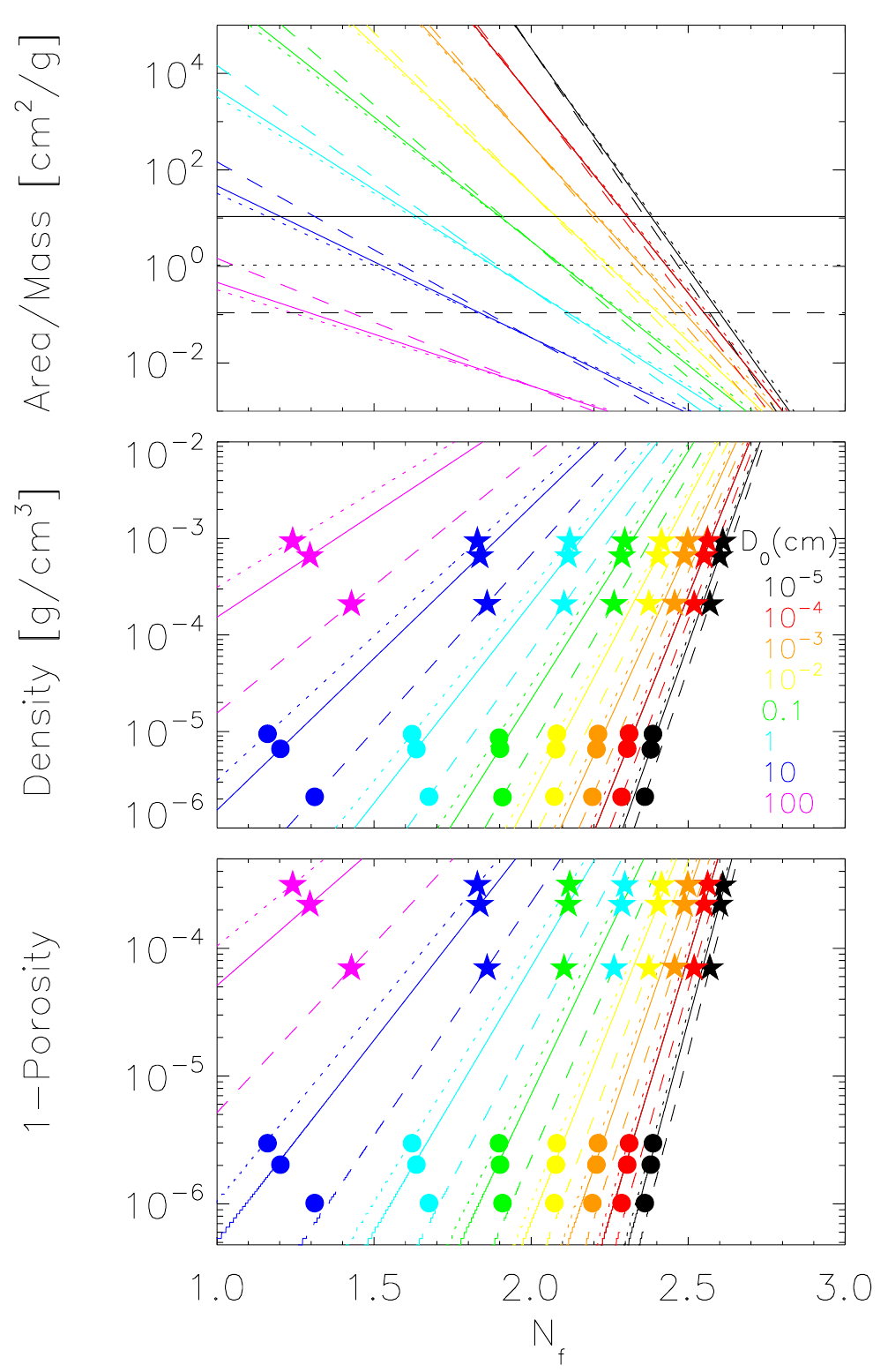}
\end{center}
\caption{The top, middle, and bottom panels show the theoretical area-to-mass ratio, density, and porosity of a body with a mass fractal structure of fractal dimension $N_f$ consisting of an aggregate of primary particles (calculated using Equation \eqref{a_m_fractal}, \eqref{rho_fractal}, and  \eqref{porosity_fractal}, respectively). The different colors correspond to different primary particle sizes, $D_{\rm 0}$, as labeled in the top panel. All the particles are assumed to have the same size and a bulk density of $\rho_{\rm 0}$ = 1 g/cm$^3$. The different line types correspond to three different overall sizes of the resulting aggregate, $D$, with $D$ = 140 m (solid), $D$ = 98 m (dotted), $D$ = 440 m (dashed) (encompassing the rage of possible sizes found by Trilling et al. (\citeyear{2018AJ....156..261T}) for 1I/'Oumuamua). The horizontal black solid line in the top panel corresponds to the area-to-mass ratio required if the non-gravitational acceleration inferred by Micheli et al. (\citeyear{2018Natur.559..223M}) from 1I/'Oumuamua's trajectory were entirely due to radiation pressure (using Equation \eqref{m_a_michelli} from Bialy \& Loeb \citeyear{2018ApJ...868L...1B}). The horizontal dotted and dashed black lines correspond to the area-to-mass ratio that would be required to cause 10\% and 1\% of the aforementioned non-gravitational acceleration, respectively. The circles (stars) in the middle and bottom panels indicate the (N$_f$, $\rho$) and (N$_f$, porosity) values that would lead to the area-to-mass ratios required to account for 100\% (1\%) of the observed non-gravitational acceleration. 
}
\label{figure1}
\end{figure}

Figure \ref{figure1} shows the theoretical area-to-mass ratio, bulk density and porosity of a body with a mass fractal structure as a function of its mass fractal dimension calculated using Equation \eqref{a_m_fractal}, \eqref{rho_fractal} and \eqref{porosity_fractal}, adopting the parameters ($D_0$, $\rho_0$, and $D$) described above. 

\section{Discussion}
\label{Discussion}

We now compare the area-to-mass ratio calculated above using Equation \eqref{a_m_fractal}, shown as the inclined lines in the top panel of Figure \ref{figure1}, to the value that would be required to support 1I/'Oumuamua's radiation pressure scenario (Equation \ref{m_a_michelli}), shown as the horizontal black solid line. For a given set of parameters ($D_0$, $\rho_0$, and $D$, defining each of the inclined lines), the required area-to-mass ratio is achieved at the mass fractal dimension given by the intersection of the corresponding inclined line and the horizontal black solid line. The circles in the middle and bottom panels of Figure \ref{figure1} show the bulk density and porosity of the corresponding aggregates.  As the figure shows, for $D$ = 98 m, 140 m, 440 m, these bulk densities are $\rho$ $\sim$ 9$\cdot$10$^{-6}$ g cm$^{-3}$, 7$\cdot$10$^{-6}$ g cm$^{-3}$, and 2$\cdot$10$^{-6}$ g cm$^{-3}$, respectively, all of which imply a porosity $\sim$ 1. 
If we were to assume that only 10\% (1\%) of the non-gravitational acceleration is due radiation pressure, the comparison would be made to 10\% (1\%) of the ${A \over m}$ value in Equation \eqref{m_a_michelli}, shown as the horizontal dotted (dashed), black line in the top panel of Figure \ref{figure1}.  The stars in the middle and bottom panels show the bulk density and porosity of the aggregates that would fulfill the above 1\% condition. As the figure shows, for $D$ = 98 m, 140 m, 440 m, these bulk densities are $\rho$ $\sim$ 9$\cdot$10$^{-4}$ g cm$^{-3}$, 7$\cdot$10$^{-4}$ g cm$^{-3}$, and 2$\cdot$10$^{-4}$ g cm$^{-3}$, respectively, all of which also imply a porosity $\sim$ 1, specifically 0.9997, 0.9998, and 0.99993, respectively.  


\subsection{A formation scenario}
A critical question is whether such a porous material could form naturally. 
At the microscopic level, there is observational evidence of the existence of fluffy aggregates with extremely low densities $<$ 10$^{-3}$ g cm$^{-3}$, detected by the GIADA instrument on ROSETTA and possibly also by Stardust (Fulle et al. \citeyear{2015ApJ...802L..12F}). These particles are thought to have formed in the pre-solar nebula having experienced no further processing (Fulle et al. \citeyear{2015ApJ...802L..12F}).  Additional evidence comes from experimental studies (Blum \& Schr{\"a}pler \citeyear{2004PhRvL..93k5503B}) and from numerical simulations of grain growth. The latter generally assume spherical primary particles 0.1 $\mu$m in size (expected for primitive interstellar grains) that stick to each other forming small, porous dust aggregates that grow with subsequent collisions. Numerous numerical studies have investigated the porosity evolution of these aggregates as they grow to planetesimal sizes (see Kataoka \citeyear{2017ASSL..445..143K} for a review). The first stage of the porosity evolution is characterized by hit-and-stick growth and during this stage the relative velocities of the colliding aggregates are low because they are strongly coupled to the gas and their movement is dominated by thermal Brownian motion. As a result, their collisional energies are not high enough to restructure the aggregates and their porosity rapidly increases as they grow. When considering only equal-mass grain colllisions, the resulting aggregates have a fractal structure with a fractal dimension of $\sim$2 (Suyama et al. \citeyear{2008ApJ...684.1310S}), whereas when considering only the collision of the aggregates with the smaller primary grains, the resulting aggregates are more compact because the colliding primary grains tend to fill up the voids more efficiently, resulting in a fractal dimension of $\sim$3. Because the growth of the dust grains involves both types of collisions,  the solution lies somewhere between these two extremes, depending on which process is dominant. 

The second stage of the porosity evolution is characterized by collisional compression and it begins when the collisional energy exceeds the rolling energy of the aggregate, defined as the energy required to roll a primary particle over a quarter of the circumference of the another primary particle in contact. In spite of its name, compression during  this stage is  inefficient and the porosity of the aggregate continues to increase as it grows. This is because most of the colliding energy is spent compressing the new voids that are created when two aggregates collide and stick to each other, rather than compressing the voids that were already present in the colliding aggregates. Suyama et al. (\citeyear{2008ApJ...684.1310S}) has investigated the porosity evolution of icy dust aggregates growing in laminar protoplanetary disks similar to the minimum-mass solar nebula via sequential equal-mass, head-on collisions. They found that the collisional compression stage results in fluffy aggregates  with a fractal dimension of 2.5 and extremely low densities $<$ 10$^{-4}$ g cm$^{-3}$, noting that this density would be even lower if one were to account for oblique collisions that result in elongated aggregates. Unequal-mass collisions can result in an increased density but the change is found to be small (Okuzumi et al. \citeyear{2009ApJ...707.1247O}, Suyama et al. \citeyear{2012ApJ...753..115S}).  Okuzumi et al. (\citeyear{2012ApJ...752..106O}) extended this model to study the growth of icy aggregates beyond the snowline of protoplanetary disks and found that the resulting aggregates at the end of the collisional compression stage have even lower densities of $\sim$ 10$^{-5}$ g cm$^{-3}$ for a wide range of aggregate sizes, encompassing 1I/'Oumuamua estimated size (see Figure 10 in Okuzumi et al. \citeyear{2012ApJ...752..106O}). This density is of the order of the value discussed above that is required to support a radiation pressure-driven scenario for 1I/'Oumuamua that is based on its internal structure rather than its geometry.  

Because the most primitive dust "aggregates" in the solar system are the comets and they have much higher densities ($\rho$ $\sim$ 0.1 g m$^{-3}$) and lower porosities ($\sim$60\%--70\%), to bridge this gap it has been suggested that the aggregates experiment static compression due to ram pressure from the disk gas and due to self-gravity (Okuzumi et al. \citeyear{2012ApJ...752..106O}), the latter becoming important only when the objects reach km-sizes. Numerical models by Kataoka et al. (\citeyear{2013A&A...557L...4K}) found that ram pressure by the disk gas can increase the density of the aggregates resulting in densities of the order of 10$^{-4}$--10$^{-3}$ g cm$^{-3}$ for aggregates smaller than a few hundred meters, encompassing 1I/'Oumuamua's estimated size. These numerical models, however, assume compressive strengths derived numerically, as the experimental studies that investigate static compaction focus on more compact grains rather than fluffy aggregates. 

There is therefore the interesting possibility that the collisional grow of icy\footnote{Silicate grains, on the other hand, would result in more efficient collisional compression because of their lower surface energy compared to icy grains, resulting in higher densities.} dust particles beyond the snowline of a protoplanetary disk might naturally produce fractal aggregates of the size of 1I/'Oumuamua with bulk densities that are low enough to support, or to contribute significantly, to the radiation pressure-driven scenario described in this paper (see the inclined black lines in Figure 1 corresponding to $D_0$ = 0.1 $\mu$m). This would not only help to explain 1I/'Oumuamua's non-gravitational acceleration, but might also shed light on its unusual physical properties (because such a fluffy aggregate would be very different from the more compact solar system objects taken as reference), and its high inferred number density (because a nearby, protoplanetary disk origin may imply that the object is not representative of an isotropic population, as suggested by Gaidos et al. \citeyear{2017RNAAS...1a..13G} and Moro-Mart\'{\i}n \citeyear{2018ApJ...866..131M} and \citeyear{2019AJ}). 

This is a hypothesis worth investigating because, if 1I/'Oumuamua were to have such origin, it would have truly opened a new observational window to study the building blocks of planets around other stars (generally limited to the two extremes of the size distribution: the dust and the planets), and this can set unprecedented constraints on planet formation models. For example, numerical models find that fluffy icy aggregates, like the ones discussed above that 1I/'Oumuamua may represent, can accelerate planetesimal growth because of their increased cross-section, helping to avoid several growth barriers (Suyama et al. \citeyear{2008ApJ...684.1310S}). They can overcome the radial drift barrier within 10 AU for a minimum mass solar nebula model, facilitating planetesimal growth in the inner regions of protoplanetary disks, outside the water snowline (Okuzumi et al. \citeyear{2012ApJ...752..106O}). They can also overcome the  fragmentation barrier if they are constituted by primary particles 0.1 $\mu$m in size because the expected maximum collisional velocities in the disk midplane are generally smaller than the fragmentation threshold velocities for these type of aggregates (Kataoka et al. \citeyear{2013A&A...557L...4K}). Finally, fluffy icy aggregates are not subject to the bouncing barrier because of the small  number of primary particles that are in contact with each other. The existence of fluffy aggregates can also have an impact on planet formation because their porosity could delay the onset of runaway growth (as the escape velocity decreases with increasing porosity, Okuzumi et al. \citeyear{2012ApJ...752..106O}).  

\subsection{Open questions}
Aspects that need to be investigated to assess the viability of the icy fractal aggregate hypothesis proposed here are the ejection mechanism from the birth protoplanetary disk, the optical properties of the aggregates (to compare to observations), and how the aggregate would be affected by a long interstellar journey. Regarding the latter, the main stresses would likely come from tidal disruption during ejection, and collisions with interstellar grains and rotational spin-up during travel (as the object would be at the top of the Relative Tensile Strength-Porosity parameter space described by Richardson et al. \citeyear{2002aste.book..501R} in their Figure 1).  Comprehensive numerical models that investigate the effect of long-term stresses on such porous aggregates are yet to be developed (Richardson et al. \citeyear{2002aste.book..501R}). Hydrocode simulations of hypervelocity impacts into asteroid-type aggregates with much larger densities showed that a porous structure damps impact energy very efficiently, protecting the integrity of the aggregate (Richardson et al. \citeyear{2002aste.book..501R}). N-body simulations that have investigated the effect of unequal-size collisions highly porous aggregates (like the ones discussed here) are limited to small mass ratios (Okuzumi et al. \citeyear{2009ApJ...707.1247O} Figure 6). These simulations found that the unequal-size collisions produce small changes in the fractal dimension of the aggregate ($\sim$ 0.1 for a mass-ratio of 10, and $\sim$0.3 for a mass-ratio of 100), but these simulations would need to be extended to include much larger mass ratios representing the collisions with the interstellar grains. Given the results of Kataoka et al. (\citeyear{2013A&A...557L...4K}) regarding aggregate compression due to ram pressure by the disk gas, another aspect that needs to be studied is whether its extremely low density could be maintained while in the parent system, during its long interstellar journey, and when entering the solar system.

\section{Conclusion}
\label{Conclusion}

1I/'Oumuamua is a known interstellar interloper exhibiting a non-gravitational acceleration in its outbound orbit that cannot be accounted for by outgassing, given its lack of cometary activity and the strict upper limits to outgassing revealed by Spitzer observations (unless the relative abundances of the common volatiles are very different from those found in comets). It has therefore been suggested that, rather than outgassing, the non-gravitational acceleration could be due to radiation pressure (Bialy \& Loeb \citeyear{2018ApJ...868L...1B}). The required area-to-mass ratio would correspond to the physical properties of a thin sheet 0.3--0.9 mm in width that would have been produced by an unknown natural or artificial processes. As an alternative to this planar-sheet scenario, we assess whether a naturally produced, mass fractal structure with a high area-to- mass ratio could account for or contribute significantly to the observed non-gravitational acceleration observed for 1I/'Omuamua. We find that the required type of aggregate would have to be extraordinarily porous, with a density of $\sim$ 10$^{-5}$ g cm$^{-3}$. Such porous aggregates can naturally arise from the collisional grow of icy dust particles beyond the snowline of a protoplanetary disk, and we propose that 1I/'Oumuamua might be one of those ejected icy aggregates. There are many open questions that need to be addressed in order to assess the viability of this scenario, including how the aggregate would be affected by ram pressure from the gas (encountered in its parent system and during travel), by tidal disruption during ejection, and by collisions with interstellar grains and rotational spin-up during travel. This hypothesis is worth investigating because if 1I/'Oumuamua were to have such an origin, its discovery could open a new observational window on to study the of the building blocks of planets around other stars (generally limited to the two extremes of size distribution, the dust and the planets), and this can set unprecedented constraints on planet formation models.

A. M.-M. thanks the anonymous referee for very helpful suggestions that have significantly improved the manuscript. After this paper was in its final form, we learned about the work of Z. Sekanina \citeyear{2019arXiv190108704S} that proposes 1I/'Oumuamua is the porous debris resulting from the disintegration before perihelion of an interstellar comet.

\end{document}